# Siloxane crosslinks with dynamic bond exchange enable shape programming in liquid-crystalline elastomers

Mohand O. Saed, Eugene M. Terentjev

Cavendish Laboratory, University of Cambridge, Cambridge CB3 0HE, U. K.

**ABSTRACT:** Liquid crystalline elastomers (LCEs) undergo reversible shape changes in response to stimuli may enable a wide range of smart applications, such as soft robot, adhesive systems or implantable medical devices. In this manuscript, we introduce new dynamic covalent chemistry based on siloxane equilibrium exchange into the LCEs to enable processing (director alignment, remolding, and welding). Unlike, the traditional siloxane based LCEs, which are produced by a other reaction schemes with irreversible bonds (e.g. hydrosilylation), here, we use a much more robust reaction (thiol-acrylate/thiol-ene 'double-click' chemistry) to obtain highly uniform dynamically crosslinked networks. Combining the siloxane crosslinker with click chemistry produces LCEs with tunable properties, low glass transition (-30°C), controllable nematic to isotropic transition (32 to 75°C), and a very high vitrification temperature (250°C). Accordingly, this class of dynamically crosslinked LCEs shows unprecedented thermal stability within the working temperature range (-50 to 140 °C), over many thermal actuation cycles without any creep. Finally, multiple LCEs sharing the same siloxane exchangeable bonds can be welded into single structures to allow for materials that sequentially and reversibly undergo multiple phase transformations in different sections of the sample.

## INTRODUCTION

Liquid crystalline elastomers (LCE) have fascinated researchers for over 30 years, since the pioneering work has laid out the foundations of theoretical understanding[1-3] and the materials chemistry.[4] The concept of a thermally-driven reversible actuation in LCE (artificial muscle)[5,6] has been at the front of the search for practical applications,[7,8] ranging from sensors[9] to soft robotics[10] (although the unique 'soft elasticity' of LCE[11] could promise an alternative route to applications in damping[12] and adhesive systems[13]).

Traditionally, the principal methodology to prepare LCE actuators has been through the hydrosilylation reaction of siloxane monomers (spacer and/or crosslinker) and vinyl or acrylate mesogens.[4,5] The alignment of an LCE by uniaxial stress[14] (often called the polydomain-monodomain transition[15]) and the method of two-step crosslinking[16] to produce the permanently aligned (monodomain) LCE capable of actuation made the foundation of the field.[17]

The original work on LCEs used siloxane-based elastomers as a material strategy due to their incredible properties (high failure strain, low glass transitions, and low moduli), attributed to the exceptionally flexible Si–O–Si linkages within the polymer backbone. However, it has proven to be problematic to achieve any useful configuration except the uniaxial alignment in flat film, due to the unavoidable limitations of two competing processes: orientation alignment and network crosslinking.[16]

Recently, the pioneering work introduced the concept of 'vitrimers' (polymer networks covalently crosslinked by a bond-exchange reaction).[18] This appears to be a useful strategy to process LCE as well (i.e. solving the alignment problem).[19] Vitrimers are much more stable than typical transient elastomer networks, allow full thermal re-molding (making the material renewable), and molding of complex shapes with intricate local alignment (which is impossible in traditional elastomers).

The first examples of such 'exchangeable LCE' (xLCE)[19] were based on the transesterification bond-exchange reaction (BER), following the original work of Leibler et al.[18] In the past few years, a number of strategies based on dynamic covalent bonds to achieve complex alignment in LCEs have been followed, such as disulfide,[20] free-radical addition fragmentation chain transfer,[21,22] exchangeable urethane bonds,[23] and more recently – the boronic transesterification.[24] However, all of these approaches are based on hydrocarbon elastomers and share the problem of continuous creep at a relatively low temperature.

Network plasticity in hydrocarbon elastomers with C-C bonds can mainly be obtained in two ways: either through associative reactions (e.g. transesterification), where the network can alter its topology while upholding the constant number of covalent bonds, or via the bond cleavage and subsequent re-forming[25] (i.e. dissociative reactions) in covalent adaptable networks (CAN). Examples of dissociative CAN reactions include Diels-Alder reaction,[26] or disulfide metathesis.[27,28].

Silicone-based elastomers (replacing carbon with silicone which results in lowering of the glass transition) are an important class of polymers, widely used in low-temperature environment, also utilized as sealants, and in microfluidic fabrications due to their extreme hydrophobic nature and ideal mechanical properties. In principle, all of the methods used to add network plasticity to hydrocarbon-based elastomers can be incorporated in silicone-based elastomers as well. For example, vinylogous urethane exchange,[29] transesterification,[30,31] boroxine bonds,[32] or Meldrum's acid-derived bonds[33] were used with crosslinked polydimethylsiloxane (PDMS) systems. However, there are certain types of dynamic exchange unique to silicone-based bonds, such as the equilibrium

exchange,[34–36] and silyl ether metathesis in the siloxane adaptable networks.[37,38]

Here we aim to 'reinvent' the use of siloxane in LCE systems, utilizing a much more robust reaction (thiol-ene 'click' chemistry) to produce highly uniform siloxane crosslinked networks.[39–41] Firstly, we employ a two-stage, one-pot, thiol-acrylate/thiol-ene 'double-click' strategy to produce LCE with tunable thermomechanical properties, see Scheme 1. Secondly, we use the adaptable topology of a network with exchangeable siloxane bonds to impart the plastic flow and processability (i.e. the ability to re-program the alignment, re-mold, and re-shape the material, see Scheme 2).

Presently, there is a good theoretical grasp of mechanics and stress relaxation in transient elastic networks. One of the important results is that the basic stress relaxation does not strongly depend on the rate of network re-crosslinking, and is mostly determined by the rate of bond breaking.[42] This allows us to study the rates of reactions in different circumstances, and extract the activation energies for breaking of these bonds. Importantly, theoretical predictions distinguish the nature of the bond-exchange mechanism. In many physically bonded networks (held together by, e.g., hydrophobic interactions, or π-π interaction in metal–ligand compounds) the effective bond strength is a random function of local microstructure. In contrast, in a covalent BER we expect a specific activation energy value $\Delta G$ to be present, and the rate of the reaction given by the activation law $\beta = \omega_0 e^{-\Delta G/k_B T}$ with the 'rate of attempts' $\omega_0$ defined by the fraction of exchangeable bonds and their relative position in the network. The stress relaxation after an instantaneous strain step $\Delta\varepsilon$ is then given by a simple exponential: $\sigma(t) = E\Delta\varepsilon \cdot e^{-\beta t}$, with $E$ the Young modulus of the elastic network (or a weakly stretched exponential if there is a variation of bond environments leading to a dispersion in the single $\beta$ value). This is in stark contrast with thermoplastic elastomers, where the distribution of relaxation times leads to a highly stretched exponential, or power-law relaxation.[43,44]

Here we study these reaction rates, by varying concentration of siloxane crosslinks in the network (affecting $\omega_0$), and varying catalyst content (which affects $\Delta G$). According to previous reports,[34–36] any base or acid catalyst can be used to trigger the siloxane bonds exchange at an elevated temperature. In this study, we select triethylamine (TEA) and tetramethlammonium siloxanolate (TMA-Si) to serve as a catalyst for the thiol-acrylate Michael addition, and subsequently for the siloxane bond exchange. Unlike TEA, TMA-Si has a much higher evaporation temperature and so it is expected to remain in the network even after prolonged exposure to high temperatures; it also offers a different (additional) route for siloxane bond exchange, see Scheme 2.

We identify the optimal processing conditions for the network alignment and plastic re-molding. These conditions serve as a guide to prepare programmed (permanently aligned, monodomain) elastomers, in a basic film and in complex shapes, and demonstrate the reversible thermal actuation of these new materials.

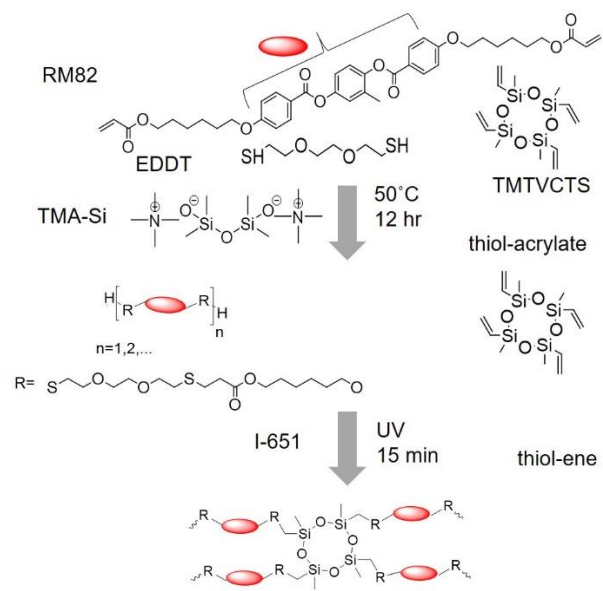

SCHEME 1. The summary of 'double-click' chemistry: the mesogenic di-acrylate (RM82) first reacts with di-thiol (EDDT), which is in excess. Secondly, the thiol-terminated oligomer chains are photo-polymerized with the vinyl bonds of the ring-siloxane, leading to the permanent network with 4-functional crosslinks.

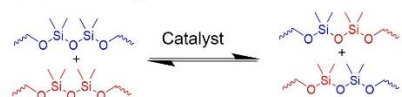
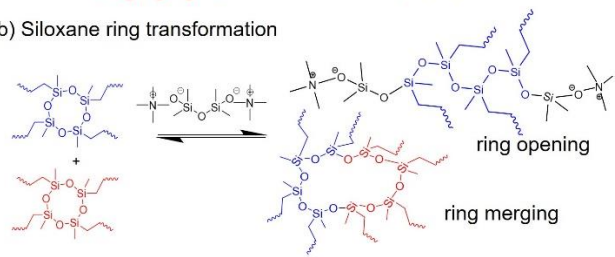

SCHEME 2. (a) The general scheme of siloxane exchange enabled by acid or base catalyst. (b) Two possible routes of siloxane exchange in this work: the siloxanolate catalyst breaks the ring and terminates the linear 4-functional siloxane crosslink, or two ring-crosslinks join into a single 8-functional ring (which may later exchange into two different 4-crosslinks due to its flexibility).

## RESULTS AND DISCUSSION

The main point of this paper is to introduce the new chemistry approach: utilizing the 'click chemistry' based on thiols, employing thiols and siloxanes to control the glass transition, and benefitting from the equilibrium siloxane exchange to impart the dynamic adaptability to the resulting elastomers. There are broad opportunities in this approach: one is free to choose a number of different units from the existing library of di-acrylate reacting monomers,[45] and di-thiol chain extenders,[46] thus achieving different LC phases and properties for various practical aims and applications.[47,48]

TABLE 1. Chemical formulations and monomer mass ratio of xLCE systems used. All systems are synthesized using thiol-acrylate/thiol-ene double click reaction. The shaded composition (40%) is what we selected for the detailed study of alignment and actuation.

| Network description | Mass of RM82 | Mass of EDDT | Mass of TMTVCTS |
|---|---|---|---|
| 20% TMTVCTS | 1 | 0.3341 | 0.0511 |
| 40% TMTVCTS | 1 | 0.3898 | 0.1023 |
| 60% TMTVCTS | 1 | 0.4454 | 0.1534 |
| 80% TMTVCTS | 1 | 0.5011 | 0.2045 |
| 100% TMTVCTS | 1 | 0.5568 | 0.2557 |

The details of our network formulations are given in the Experimental Section. Here it is important to introduce the notation: we characterize the material composition by the mol fraction of reacting bonds, thiol-acrylate and thiol-vinyl, always taking the content of mesogenic di-acrylate RM82 monomer as 100% (or 1 molar ratio). Then our lowest crosslinking density network, labelled as "20% crosslinked" has 20% (or 0.2 molar ratio) of vinyl bonds on 4-functional ring-siloxane crosslinks, and accordingly, the stoichiometric amount of 120% (or 1.2 molar ratio) of thiols on the di-functional chain extender EDDT, see Table 1. At the opposite end we have the highly crosslinked network, labelled as "100% crosslinked", which has 100% vinyl bonds (1:1 with acrylate bonds of the mesogens), and accordingly 200% (or 2 molar ratio) of thiols. For instance, according to this nomenclature, the "100% crosslinked" network has exactly two RM82 mesogens per crosslink, that is, on average network strands contain just one RM82 rod between two thiols. In the same way, the "20% network" has its strands, on average, with 5 RM82 rods separated by thiol spacers.

The DSC results of a series of materials enabled by varying molar ratio of ring-siloxane crosslinks are shown in Figure 1. The glass transition ($T_g$) is around -30°C with very little changes even when the crosslinking density is significantly increased, which can be attributed to flexibility of the siloxane crosslinker and the associated reduction of rigid mesogenic units. On the other hand, the reduction of these mesogenic units reduces the nematic-isotropic transition ($T_{ni}$). Note, that even the "100% crosslinked" LCE has a broad range of the liquid-crystalline phase below $T_{ni}$~32C. The X-ray proof that this phase is nematic is given later, in Figure 5(c).

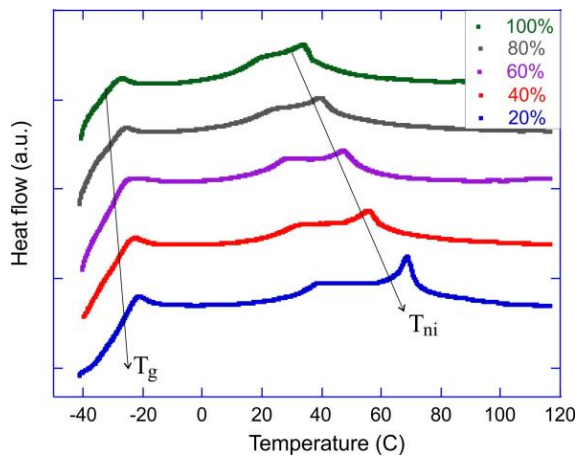

FIGURE 1. Differential scanning calorimetry (DSC) of our networks, on heating, with the different crosslinking density labelled on the plot, showing the glass- and the nematic-isotropic transition temperature variation with composition.

Figure 2 shows the results of a typical stress-relaxation in the siloxane-based xLCE, which takes place after an instant fixed-strain is imposed on the sample (maintaining the constant temperature). The results are presented via a scaled relaxation function $\sigma(t)/\sigma_{max}$, in order to focus purely on the time dependence. The normalized stress as function of time for 40% TMTVCTS samples containing various TEA and TMA-Si concentrations is shown in Figure 2(a). We compare the networks a total of 1wt% of a catalyst mixture of TMA-Si and TEA in ratios 1:0, 0.3:0.7, 0.1:0.9 and 0:1, respectively. The slowest relaxation is in the 1% TEA sample (labelled as 0% TMA-Si in the plot); the increasing fraction of TMA-Si makes the bond exchange faster. Both of these amines can trigger the relaxation of the siloxane elastomer, however, TEA is more volatile catalyst at elevated temperature. Therefore, it has slower the stress relaxation compared to TMA-Si. For comparison, we also show the relaxation of an xLCE with 3wt% of TMA-Si catalyst, which predictably is much faster.

The fitting of such scaled stress relaxation curves with the basic exponential relaxation, for 1% TMA-Si is shown in Figure 2(b). It gives the characteristic relaxation time $\tau$ for each material and temperature. As expected, increasing the temperature accelerated the relaxation, where at 210°C the elastomer is fully relaxed after 7000s due to its internal plastic flow. To study the influence of the siloxane concentration on the stress relaxation, we tested siloxane crosslinked networks containing various siloxane concentrations (e.g. 20, 40, and 100 functional mol %), each network having the same amount of catalyst (1 wt% of TMA-Si). The data of the relaxation times for various samples were then collated at different temperatures to generate the Arrhenius plot (Figure 3). That is, we plot τ(T) on the logarithmic scale, and fit the data with the activation law $\ln[\tau] = \text{const} + \Delta G/k_B T$.

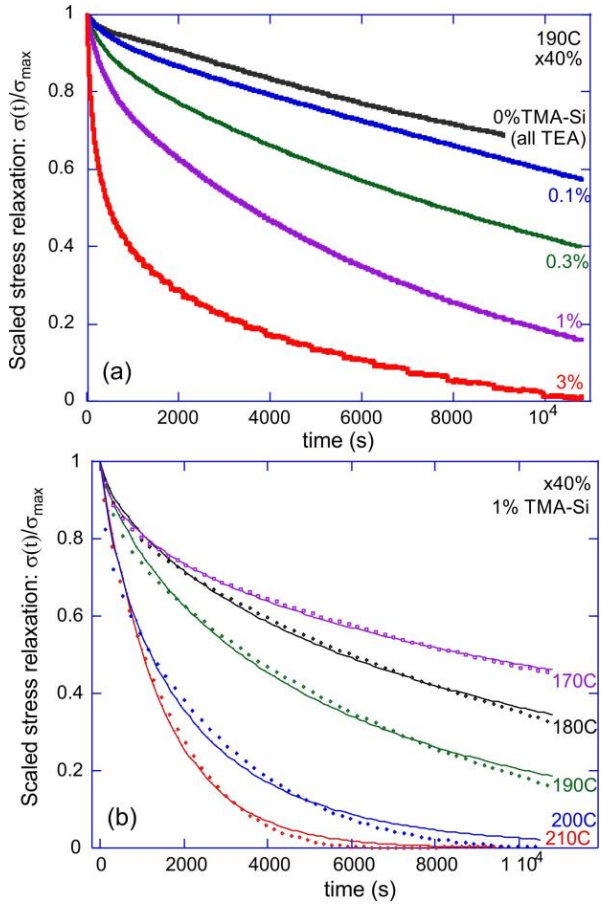

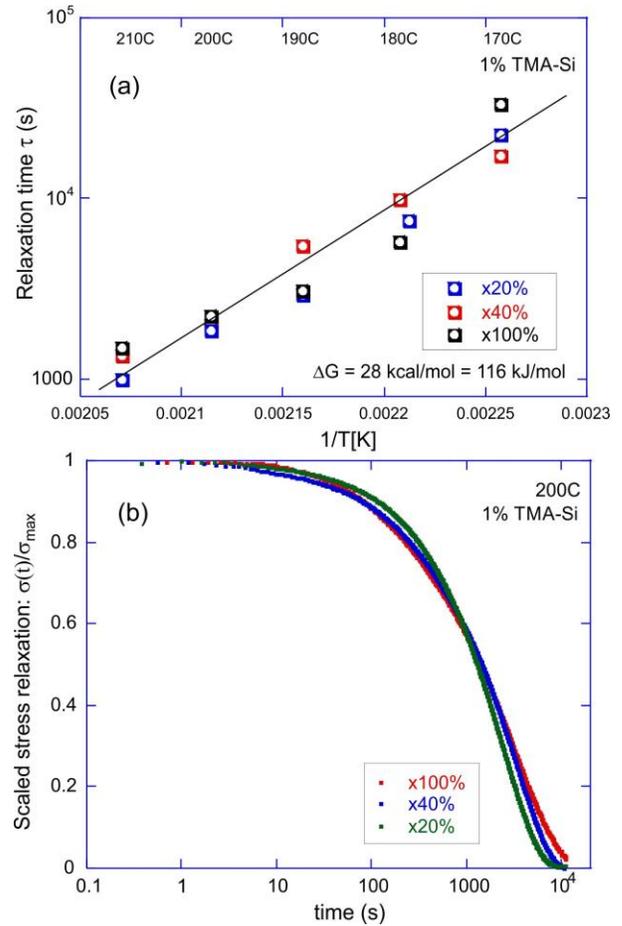

FIGURE 2. (a) Scaled stress-relaxation $\sigma(t)/\sigma_{max}$ for the 40%-crosslinked xLCE at T=190°C, and several concentrations of catalyst, as labelled in the plot. (b) Stress relaxation curves for with 1 wt% of TMA-Si catalyst, at several temperatures labelled in the plot. Dashed lines are the fits with exponential function, which produce the relaxation time $\tau = 1/\beta$.

FIGURE 3. (a) The Arrhenius plots for the relaxation time $\tau(T)$, for different xLCE networks. The slope of the linear fitting gives the bond strength $\Delta G \approx 28$ kcal/mol, and the additive constant gives the 'rate of attempts' $\omega_0$. Three data sets are for the 20%, 40% and 100% crosslinked networks, all fit with the same activation law. (b) Comparison of the scaled stress relaxation at 200C for the networks with 20%, 40% and 100% crosslinking density. The logarithmic time axis helps the comparison at short and long times.

It is clear that data sets show a single value of activation energy $\Delta G \approx 28$ kcal/mol (or 116 kJ/mol), which corresponds to about 45 $k_B T$ at room temperature, and is in good agreement with the results of *Xie et al.*[49] who used 0.1 wt% of sodium octanoate as catalyst in a much higher siloxane concentration elastomer (Sylgard 184 PDMS). In comparison, in the work of Leibler et al.[18], the transesterification with the zinc acetate catalyst had the activation energy $\Delta G \approx 20$ kcal/mol (or 34 $k_B T$). It is expected, and reassuring, that the single value of activation energy $\Delta G$ describes the macroscopic stress relaxation: this is a clear signature of a distinct reaction pathway, in our case depicted in Scheme 2.

Surprisingly, siloxane elastomers with very different concentration of crosslinker appear to have the same 'rate of attempts' $\omega_0$ in their relaxation behavior. It is confirmed by comparing the relaxation curves themselves for these different networks at the same temperature in Figure 3(b). To explain this, we argue that the first exchange route in Scheme 2 (the ring opening) is the dominant process, or perhaps even the only possible route for the bond exchange. We used a relatively large amount of catalyst in this system ~1 wt% (i.e. 5 times more than in McCarthy et al.[36] (same catalyst) and 10 times more than in Xie et al.[49] Catalyst helps terminate the rings after their opening. Besides, we believe the contact of two siloxane-ring crossliners in the stretched network has a low probability, while the mobile TMA-Si can reach any location in the network. As the catalyst content was the same in the data shown in Figure 3, so are the relaxation rates.

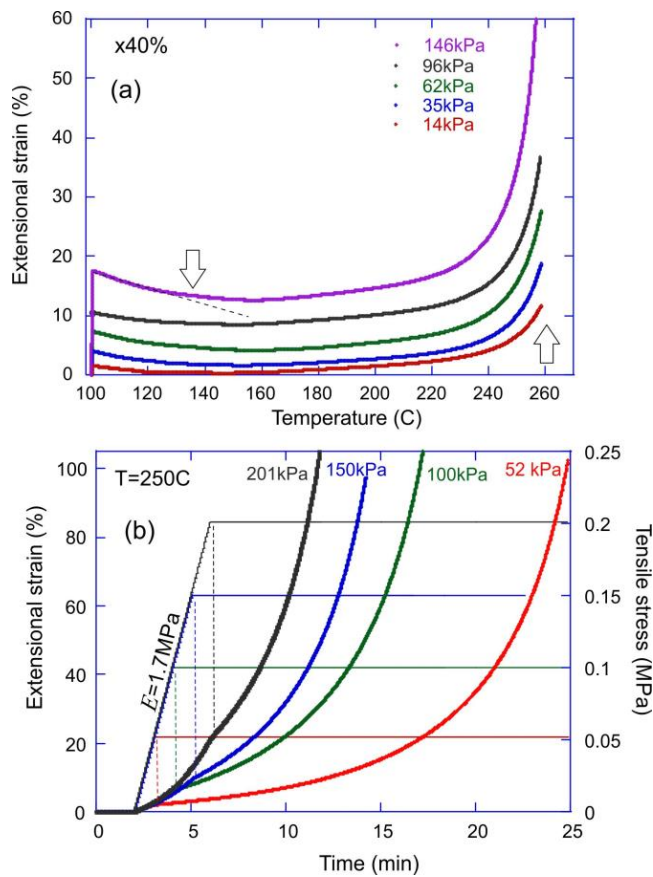

in the isotropic phase (the 40% crosslinked xLCE is chosen for this demonstration). First, we apply a given stress (as labelled in the plot), and register the resulting extensional strain, which gives the value of the Young modulus of the material (E ≈ 880 kPa in the 40% crosslinked xLCE), and then increase the temperature at a constant rate of 2°/min.

The first observation one makes is the classical rubber-elastic response: as the entropic rubber modulus increases with temperature, at constant stress the strain decreases.[50] However, as the temperature increases further, and the bond-exchange becomes more prominent, the plastic flow (creep) starts being noticeable. The region where the data deviates from the initial rubber-elastic decreasing slope is identified as the transition to plastic flow, the vitrification point $T_v$: apparently it does not depend on the applied stress.[51] One should expect some creep under stress in network with siloxane-exchange above 140-150C, although the rapid flow only sets in at a much higher temperature (over 250C).

We use this regime of stress-induced plastic flow to program our xLCE materials into monodomain aligned state. Figure 4(b) illustrates the process: we bring the sample to a high temperature (T=250C) as suggested by the results of iso-stress test, apply a constant tensile stress to a level labelled in the plot, and then keep the sample at these constant temperature and stress until its elongation reaches 100% (clearly, this happens faster at higher stress, but in all cases the process takes several minutes and allows easy control). We deem the 100% elongation to be sufficient to impart the fully uniaxial monodomain alignment to our xLCE, and then take the programmed sample off the stress and heating. Figure 5 illustrates the aligned sample (comparing it with the initial polydomain xLCE in Fig.5a). This programmed alignment is permanent as long as we do not allow the sample temperature to raise above 140C, cf. Fig. 4(a), when the residual creep would cause a gradual loss of alignment (increasing at even higher temperatures). However, as with all exchangeable xLCE systems, we can re-program the material to a different shape and state of alignment by a subsequent process.

FIGURE 4. (a) The 'dilatometry' curves strain changing with temperature at constant stress (labelled on the plot). The equilibrium rubber modulus of the 40% crosslinked xLCE at 100C (in the isotropic phase) is E = 880 kPa, which gives the initial strain in all curves. At temperatures below $T_v$ we see the classical effect of entropic elasticity, the rubber modulus linearly increasing with T. The onset of plastic flow occurs around 140-160C, with the network in free-flow above 250C. (b) Programming of the aligned monodomain in xLCE. With a sample kept at constant T=250C, we apply a tensile stress (labelled on the plot), and allow the plastic flow to reach 100% extension, at which point we consider the structure aligned, and take the sample out of the device. The time for this programming is several minutes, depending on the applied tensile load.

Figure 4 presents the results of the dynamic response of our exchangeable elastomers, due to the siloxane exchange reaction allowing plastic flow under stress, at a sufficiently high temperature. First, we study the 'iso-stress' response on changing temperature, which is often incorrectly called 'dilatometry' in the literature. Dilatometry is a valid test of changing the sample volume, while in these elastomer networks the volume is certainly remaining constant. What we test is how the extension changes with temperature in the sample under constant tensile stress, Figure 4(a). It is not an easy experiment in a LCE material because the effect of LCE thermal actuation produces a massive strain change on heating into the isotropic phase. Here we are concerned with the elastic-plastic transition of the exchangeable network, and so we start at T=100C, well

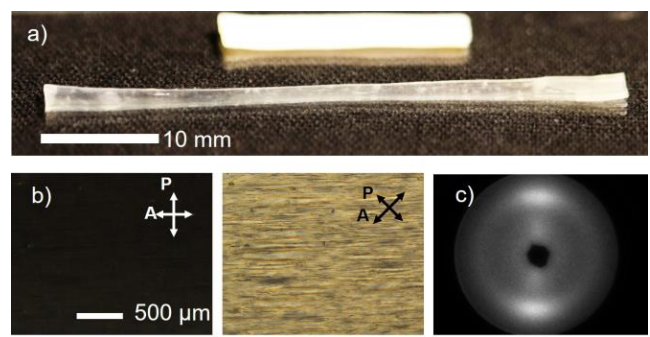

FIGURE 5. (a) The initial polydomain and uniaxially aligned monodomain xLCE, programmed by its plastic flow to 100% elongation. (b) The microscopy images between crossed polars indications alignment. (c) X-ray image confirming a good uniaxial nematic alignment, also letting us calculate the order parameter Q=0.62

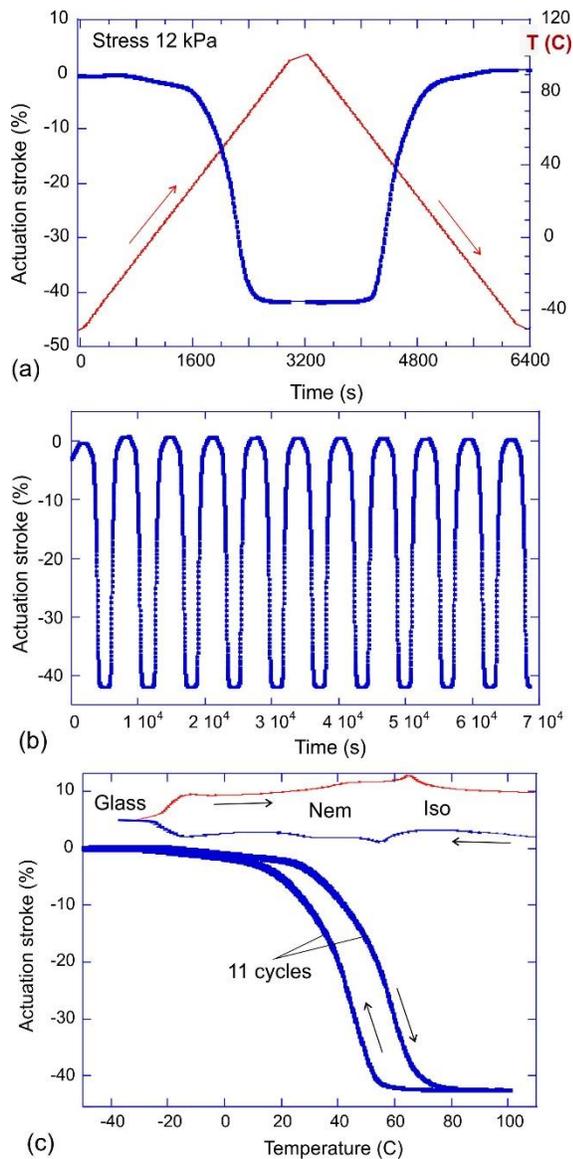

FIGURE 6. (a) One cycle of heating-cooling of the xLCE with 40% crosslinks, with the temperature on the right y-axis, and the associated sample strain showing the classical reversible thermal actuation of LCE. (b) The cyclic contraction-extension during 11 heating cycles. (c) The actuation strain plotted against temperature, showing the reproducibility of actuation, and also the extent of thermal hysteresis at the heating rate of 3°/min applied in this test.

Having programmed the uniaxial monodomain alignment in xLCE, we can now examine its actuation response to reversible heating and cooling through the nematic-isotropic transition. Figure 6 illustrates different elements of this test, carried out in a DMA instrument under a low constant stress (of 12 kPa) to ensure the sample is straight and taut. In Figure 6(a) we zoom-in on one cycle of heating and cooling, over the range of -50C to 90C (the 40% xLCE is used, with its $T_g \approx$ -20C and $T_{ni} \approx$ 60C). The sample starts rapid contraction when the temperatures approaches 30C, and reaches the saturation strain of over 40% at around 70C (both values are clearly affected by the dynamics of temperature change). On cooling the cycle reverses, and Figure 6(b) illustrates the remarkable stability of this spontaneous contraction-expansion over 11 cycles of temperature. We expect no creep of thermal degradation to occur in our xLCE materials as the temperature never reached the levels where plastic creep might set in. The same 11 cycles of heating and cooling are shown in Figure 6(c) as actuation strain against temperature: all heating and all cooling strokes are on top of each other, however, we also notice a clear hysteresis of the nematic-isotropic transition. To support this observation, Figure 6(c) also shows the DSC scans (scaled, in a.u.) on heating and cooling, at the top of the plot, to illustrate where the glass and nematic transitions are in each direction.

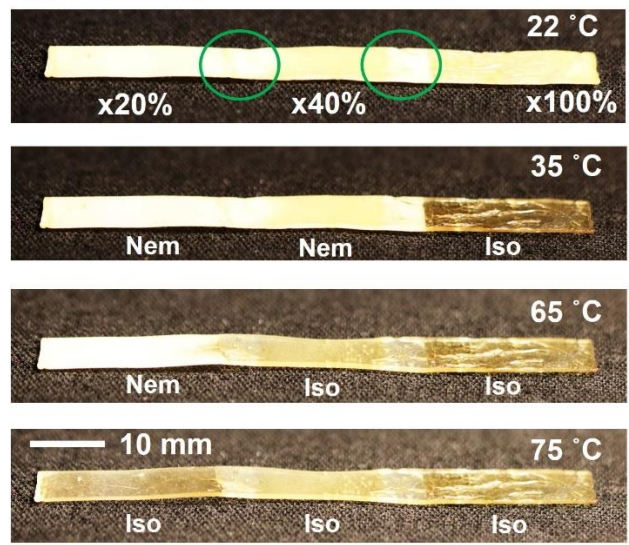

FIGURE 7. The thermally molded continuous strip combining three different xLCE materials: the 20%, 40%, and 100% crosslinked, as listed in Table 1. Since the nematic-isotropic transition in each material is at a different temperature, we see different sections of the sample become sequentially isotropic on heating.

The wide separation of the nematic transition and the vitrification temperature, at which the plastic creep starts to occur in the xLCE under stress is the reason for stability of the thermal actuation, and the programmed alignment pattern. The second remarkable aspect of xLCE (and indeed all vitrimers) is their capacity of thermal molding. We demonstrate this capacity on the following example: we take three different xLCE materials: with 20%, 40%, and 100% crosslinking density (cf. Table 1, and Figure 1). Three strips are molded together into one continuous sample, at T=250C and high pressure, kept overnight. It is reassuring that the thiol-siloxane mesogenic system has such a remarkable thermal stability (few polymers will withstand several hours at 250C without any degradation). Figure 7 illustrates the result of this molding, where one cannot distinguish the initial overlap regions. In this photo, at room temperature, all three sections are in the polydomain nematic state, and so white (strongly scattering light). Then, on heating this strip, we see the sequential phase transitions into the isotropic phase that take place in different sections of the otherwise continuous polymer strip: first the 100% crosslinked section becomes isotropic (transparent, no longer scattering light), then the 40% section, until finally the whole strip becomes isotropic.

This paper is presenting the new material concept, and its characterization, so we did not aim to construct complex actuating shapes – merely to demonstrate the capacity to mold together different xLCE materials containing exchangeable siloxane bonds and the appropriate catalyst. This offers rich design options for complicated actuating shapes and constructions for practical applications.

**CONCLUSIONS:**

In summary, we have designed a new class of exchangeable LCE network using a robust click chemistry ('double click' of thiol-acrylate and thiol-ene) and utilized the siloxane segments in the crosslinkers. These materials have several important advantages over the previous generations of LCE, which also use the commercial off-the-shelf starting ingredients: [1] the presence of thiols and siloxanes makes the glass transition naturally low; [2] they allow good control of the nematic transition, including bringing the $T_{ni}$ down to the 'human range' of 30-40oC that allows control of actuation by body heat; [3] the siloxane bond-exchange reaction imparts the dynamic network properties, similar to vitrimers: the plastic flow under stress at a high temperature allows both the programming of monodomain textures in the xLCE, and the (re)molding of plastic samples into desired structures. Obviously, the use of siloxane in the LCE is not limited to the crosslinkers, as in this paper: main-chain LCE with siloxane spacers, and side-chain LCE with siloxane backbone are well known. In these systems, like in our work here, the siloxane bond-exchange would be equally possible and useful.

**EXPERIMENTAL DETAILS**

*Materials:* Acrylate liquid crystal (LC) monomer, RM82 was purchased from Wilshire Technologies, Inc. Thiol chain extender, EDDT and vinyl siloxane crosslinker, 2,4,6,8-Tetramethyl-2,4,6,8-tetravinyl cyclotetrasiloxane (TMTVCTS), were purchased from Sigma-Aldrich. Tetramethulammonium siloxanolate, TMA-Si was purchased from Gelest and used as Michael addition base-catalyst and as anionic initiator to the siloxane bond exchange. Triethylamine (TEA) was purchased from Sigma-Aldrich and used as base-catalyst. The photoinitiator, Irgacure I-651, purchased from Sigma-Aldrich. Toluene and Tetrahydrofuran, THF were purchased from Sigma-Aldrich and used as solvent.

*Preparation of networks:* We took advantage of the thiol reaction selectivity towards different functional groups to design one pot two-step thiol-acrylate/thiol-ene reaction sequence to prepare LCE with controllable nematic transition temperatures from the commercially available starting materials. We first prepare LC oligomers via the self-limiting thiol-acrylate Michael addition between a mesogenic diacrylate (RM82) and an isotropic dithiol (EDDT). The Michael addition was catalyzed via TMA-Si or TEA. By controlling the molar ratio of thiol to acrylate, thiol-terminated oligomers were obtained. The di-thiol oligomer is radically crosslinked with vinyl siloxane crosslinker, TMTVCTS. The overall reaction scheme is similar a previously reported method.[48] In a 25 ml vial the intended amount of catalyst TMA-Si (0.1, 0.3, 1, or 3 wt%), was initially dissolved in a mixture of solvent (20 wt% THF and 20 wt% toluene), to this solution, RM82 was added and heated to 80°C until fully dissolving. After the mixture was cooled down to room temperature, I-651 (1.5 wt%), EDDT, and TMTVCTS were added and mixed vigorously using vortex mixer. The monomers solution was degassed using a vacuum chamber and then quickly transferred into a mold (two glass sides with 1 mm spacer coated with ran-x, anti-sticking agent). The monomer mixture was kept at 50°C to fully oligomerize via Michael addition reaction for 12 h. Then the thiol-terminated oligomer was photopolymerized with TMTVCTS via 365 nm UV light for 15 min at 50 °C. The ratio of thiol, acrylate, and vinyl molar functional groups was kept constant in all samples. The molar ratio used was 1.0 acrylate:1.4 thiol:0.4 vinyl, unless otherwise noted. After the polymerization was done, the samples were removed from the mold and placed in a vacuum oven at 80°C for 12 h to remove the solvents.

*Differential Scanning Calorimetry (DSC):* DSC4000 PerkinElmer was used to obtain the transition temperatures. Samples with ≈10 mg were loaded into standard aluminum DSC pans. The samples were heated to 120 °C at 10 °C min−1, held isothermally for 5 min to undo the thermal history, and cooled to −50 °C at 10 °C min−1. Then samples were heated again to 120 °C to obtain the data. Tg could be found at the step change in the slope of the heat flow signal and Tni could be obtained at local minimum of the endothermic peak. The sample was run three times.

*Stress Relaxation Measurements:* DMAQ800 (TA instruments) was used to characterize the relaxation behavior of siloxane crosslinked LCE. Samples with dimensions of ≈15 mm × 5 mm × 0.9 mm were tested. All of the samples were tested under constant uniaxial strain 3% imposed at t = 0, the strain was held constant isothermally for 180 min at 170, 180, 190, 200, or 210 °C. Prior imposing the strain, samples were kept at the desired temperature for 5 min. Samples were annealed at 80 °C for 12 h before the relaxation test.

*Iso-force Measurements:* DMAQ800 (TA instruments) was used to characterize the plastic flow of siloxane crosslinked LCE induced by siloxane bond exchange as a function of temperature. Samples with dimensions of ≈15 mm × 5 mm × 0.9 mm were tested. All of the samples were tested under constant uniaxial stress of 14, 35, 65, 96, or 146 kPa imposed at t = 0, the stress was held constant while the temperature was ramped at 2 °C/min until 260 °C. Prior imposing the stress, samples were kept at the desired temperature for 5 min. Samples were annealed at 80 °C for 12 h before the relaxation test.

*Programing monodomain Measurements:* DMAQ800 (TA instruments) was used to align polydomain samples into monodomain via creep test. Samples with dimensions of ≈15 mm × 5 mm × 0.9 mm were tested. All samples were tested under constant uniaxial stress of 50, 100, 150, or 200 kPa imposed at t = 0, the stress was held constant isothermally at 250 °C until the strain reached 100%. Prior imposing the stress, samples were kept at the desired temperature for 2 min. After reaching 100% strain the samples were kept starched while cooling to room temperature. Samples were annealed at 80 °C for 12 h before the relaxation test.

*Wide angle x-ray scattering (WAXS):* The phase of the monodomain LCE at room temperature was characterized using a Philips diffractometer using a Philips Copper target (PW-2233/20) with the wavelength of 0.154 nm. The beam size was ~ 0.7 x 0.7 mm² with flux of 4X 10^9 X-ray/s. The distance between the sample and the imaging area was 100 mm. The sample (0.5 mm x 6.5 mm and 20 mm) was exposed to the x-ray surce for 20 seconds.

*Actuation Measurements*: Discovery DMA850 (TA instruments) was used to measure the actuation performance for the monodomain film. Rectangular samples measuring approximately 15 mm × 5 mm × 0.5 mm were tested in tensile mode. To measure actuation strain, a constant stress (12 kPa) was applied to the LCE film; each sample was heated and cooled at least 11 times from 100 to -50 °C, at 3 °C min–1.

*Welding conditions*: Moore hydraulic press (Birmingham, England) was used to hot press the LCE samples. Samples (x20, x40, and x100) were first held at 250 ˚C for 15 min before applying a load on 0.5 ton. The samples were allowed to cool (overnight) to room temperature under the applied load.